\documentclass[%
reprint,
 amsmath,
 amssymb,
 aps,
 prx,
]{revtex4-2}

\usepackage{xcolor}
\usepackage[colorlinks=true,citecolor=blue]{hyperref}
\usepackage{graphicx}
\usepackage{subfigure}
\usepackage{bookmark}
\usepackage{hyperref}
\usepackage[normalem]{ulem}
\usepackage{dsfont} 
\usepackage{upgreek}
\usepackage{booktabs}

\newcommand{\comma}{~,}
\newcommand{\fullstop}{~.}

\newcommand{\abs}[1]{\left\vert #1 \right\vert}
\newcommand{\cabs}[1]{\vert #1 \vert}

\newcommand{\cerw}[1]{\langle #1 \rangle}
\newcommand{\komm}[2]{\left[ #1, #2 \right]}

\newcommand{\cakomm}[2]{\lbrace #1, #2 \rbrace}
\newcommand{\ket}[1]{\left\vert #1 \right\rangle}
\renewcommand{\d}{\mathrm{d}}
\newcommand{\Tr}{\mathrm{Tr}}

\newcommand{\rini}{r_0}
\newcommand{\riniopt}{r_0^\mathrm{opt}}
\newcommand{\rdec}{r}
\newcommand{\rdecopt}{r^\mathrm{opt}}
\newcommand{\Sigmadet}{\Sigma_\mathrm{det}}
\newcommand{\Sigmadetsq}{\Sigmadet^2}
\newcommand{\bfDelta}{\mathbf{\Delta}}
\newcommand{\Xidet}{\Xi_\mathrm{det}}
\newcommand{\Xidetsq}{\Xidet^2}
\newcommand{\sigmaadd}{\sigma_\mathrm{add}}
\newcommand{\sigmaaddsq}{\sigmaadd^2}
\newcommand{\sigmaaddsqopt}{\sigma_\mathrm{add,opt}^2}
\newcommand{\gammarel}{\gamma_\mathrm{rel}}
\newcommand{\gammaphi}{\gamma_\phi}
\newcommand{\oma}{\omega_\mathrm{a}}
\newcommand{\oms}{\omega_\mathrm{s}}
\newcommand{\Gammacoll}{\Gamma_\mathrm{coll}}
\newcommand{\topt}{t^\mathrm{opt}}
\newcommand{\Gopt}{G^\mathrm{opt}}
\newcommand{\Gmax}{G^\mathrm{max}}
\newcommand{\etaphi}{\eta_\phi}
\newcommand{\etarel}{\eta_\mathrm{rel}}
\newcommand{\Deltaphisqamp}{(\bfDelta \phi)^2_\mathrm{amp}}
\newcommand{\Deltaphisqampideal}{(\bfDelta \phi)^2_\mathrm{amp,ideal}}
\newcommand{\Deltaphisqproj}{(\bfDelta \phi)^2_\mathrm{proj}}
\newcommand{\Deltaphisqdet}{(\bfDelta \phi)^2_\mathrm{det}}
\newcommand{\Deltaphisqadd}{(\bfDelta \phi)^2_\mathrm{add}}
\newcommand{\DeltaphisqSQL}{(\bfDelta \phi)^2_\mathrm{SQL}}
\newcommand{\DeltaSysq}{(\bfDelta S_y)^2}
\newcommand{\DeltaSysqini}{\DeltaSysq_\mathrm{ini}}
\newcommand{\DeltaSysqfin}{\DeltaSysq_\mathrm{final}}

\newcommand{\thetitle}{Squeezed superradiance enables robust entanglement-enhanced metrology even with highly imperfect readout}
\newcommand{\theauthors}{Martin Koppenh\"ofer$^{1}$, Peter Groszkowski$^{1,2}$, and A.\ A.\ Clerk$^1$}
\newcommand{\theaffiliations}{$^1$Pritzker School of Molecular Engineering, University of Chicago, Chicago, IL 60637, USA \\
$^2$National Center for Computational Sciences, Oak Ridge National Laboratory, TN 37831, USA}
\newcommand{\prlsection}[1]{\paragraph*{#1---}}

\def\FigSsrOatVsNValueXidetsq{1}
\def\FigSsrOatVsNValuesignal{10^{-5}}
\def\FigSsrOatVsNValueetaphi{0.5}
\def\FigSsrOatVsNValueetarel{0.5}

\def\FigSsrOatVsEtaValueN{50}

\def\FigOptimizedEstimationErrorValuesignal{10^{-5}}
\def\FigOptimizedEstimationErrorValueInterpolationN{7}
\def\FigOptimizedEstimationErrorValueInterpolationXi{25}

\def\FigOptimalSqueezingGainValueN{400}

\def\FigOptimalTimeAddedNoiseValuesignal{10^{-5}}
\def\FigOptimalTimeAddedNoiseValueN{400}

\def\FigScalingEstimationErrorValueFitRangeLow{10}
\def\FigScalingEstimationErrorValueFitRangeHigh{400}

\def\FigContributionsEstimationErrorValueXidetsq{1}
\def\FigContributionsEstimationErrorValuesignal{10^{-5}}
\def\FigContributionsEstimationErrorValueN{400}

\def\FigGainAndAddedNoiseOversqueezedValueN{100}
\def\FigGainAndAddedNoiseOversqueezedValuesignal{10^{-5}}
\def\FigGainAndAddedNoiseOversqueezedValueRA{0.0}
\def\FigGainAndAddedNoiseOversqueezedValueRB{4.0}

\begin{document}

\title{\thetitle}
\author{\theauthors}
\affiliation{\theaffiliations}
\date{\today}

\begin{abstract}
Quantum metrology protocols using entangled states of large spin ensembles attempt to achieve measurement sensitivities surpassing the standard quantum limit (SQL), but in many cases they are severely limited by even small amounts of technical noise associated with imperfect sensor readout.  
Amplification strategies based on time-reversed coherent spin-squeezing dynamics have been devised to mitigate this issue, but are unfortunately very sensitive to dissipation, requiring a large single-spin cooperativity to be effective.   
Here, we propose a new dissipative protocol that combines amplification and squeezed fluctuations.  It enables the use of entangled spin states for sensing well beyond the SQL even in the presence of significant readout noise.  Further, it has a strong resilience against undesired single-spin dissipation, requiring only a large {\it collective} cooperativity to be effective.  
\end{abstract}

\maketitle


\prlsection{Introduction}

Entanglement-assisted quantum metrology protocols use nonclassical states to enhance the sensitivity of interferometric measurements of small signals \cite{Giovannetti2006,Degen2017,Pezze2018}. 
Well-known examples of such sensing states in ensembles of spin systems are spin-squeezed states and GHZ states \cite{Kitagawa1993,Bollinger1996}.
For a spin-squeezed state (SSS), the projection-noise distribution of the state is reshaped to reduce the fluctuations in the collective spin component associated with signal acquisition.
Besides this intrinsic projection noise, the final measurement of the spin ensemble can contribute additional detector noise \cite{Degen2017}, which severely limits the sensitivity improvement achievable by spin squeezing.
While this detection noise is not a fundamental limitation, it can pose a seemingly insurmountable practical challenge in many leading platforms for quantum metrology (see e.g.~\cite{Barry2020}).

To circumvent this detection-noise issue, twist-untwist protocols have been proposed \cite{Davis2016,Hosten2016}, where the SSS is transformed into a coherent-spin state (CSS) by time-reversed unitary spin-squeezing dynamics (i.e., the ``untwist'' step) prior to readout. 
These protocols can be understood in the broader context of interaction-based nonlinear readout \cite{Leibfried2004,Leibfried2005,Nolan2017} and pre-measurement control operations \cite{Len2022,Zhou2022}. 
They implement effective amplification dynamics in the spin ensemble:  the untwist step increases both the intrinsic projection noise of the ensemble as well as the non-zero polarization encoding the small signal of interest. 
The vast majority of theoretical studies and experimental demonstrations use a collective one-axis-twisting (OAT) Hamiltonian to implement the twist-untwist dynamics \cite{Davis2016,Hosten2016,Schulte2020,Chu2021,Colombo2021,FN1}.
A drawback of these \emph{unitary} protocols is their limited robustness against undesired dissipation, which raises the question whether more robust amplification dynamics that is compatible with spin squeezing could be implemented using \emph{dissipative} dynamics.

In this work, we introduce such a dissipative version of a twist-untwist amplification protocol, which allows one to use SSS for quantum sensing in the presence of readout noise. 
Surprisingly, its sensitivity surpasses the standard quantum limit (SQL) even if the detection noise is orders of magnitude larger than the projection noise of a CSS.
Moreover, our protocol is more robust against undesired dissipation than unitary protocols:  it requires only a large \emph{collective} cooperativity, whereas unitary protocols require at least a large \emph{single-spin} cooperativity, an experimentally much more demanding condition.
The dissipative amplification dynamics is caused by collective decay of a spin ensemble coupled to an effective squeezed reservoir.  
We stress that this can be engineered without having to explicitly generate squeezed light, and is compatible with a variety of experimental platforms.
While previous works have studied related setups, their focus was almost exclusively on spin squeezing in the dissipative steady state \cite{Agarwal1989,Agarwal1990,Agarwal1994,Kuzmich1997,DallaTorre2013,Borregaard2017,Groszkowski2022,GutierrezJauregui2022}.  
In contrast, our focus is not on the steady state, but we instead explicitly characterize and utilize the unstable \emph{transient} dynamics when the spins are initialized in a highly excited state.

\begin{figure}
	\centering
	\includegraphics[width=0.35\textwidth]{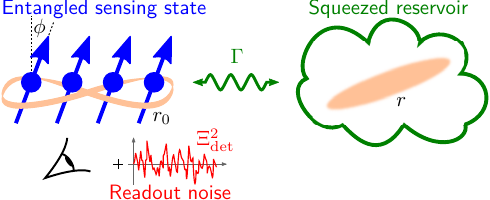}
	\caption{
		Sketch of a dissipative spin-amplification protocol with sensitivity below the standard quantum limit (SQL).
		A small parameter $\phi$ to be measured is encoded in the transverse polarization of an entangled state of $N$ spins with spin-squeezing parameter $\rini$.
		Readout of the state adds an amount $\Sigmadetsq = \Xidetsq N/4$ of detection noise. 
		The transient dynamics of squeezed superradiant (SSR) decay, i.e., collective decay to a squeezed bath with squeezing parameter $\rdec$ and decay rate $\Gamma$, reduces the detrimental impact of detection noise and allows one to reach an estimation error $\Deltaphisqamp$ below the SQL.
	}
	\label{fig:Sketch}
\end{figure}

\prlsection{Estimation error and amplification}

\begin{figure*}
	\centering
		\includegraphics[width=0.45\textwidth]{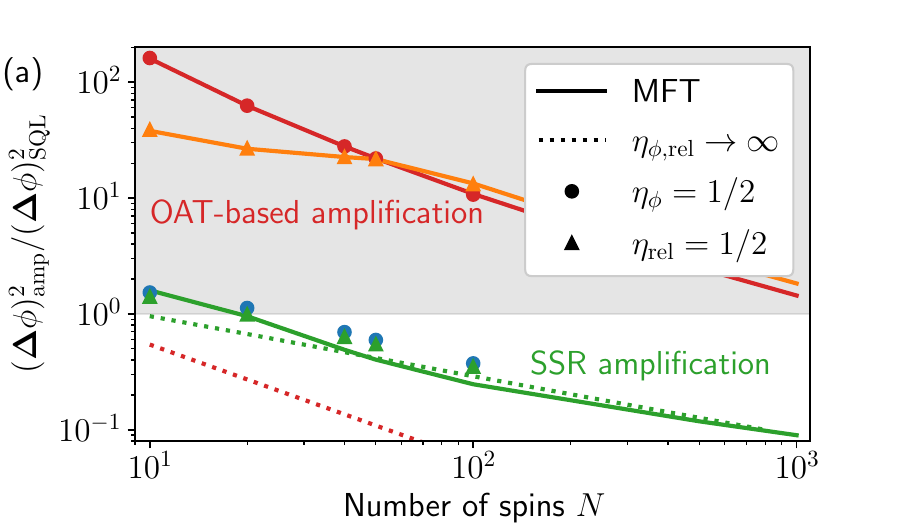}
		\includegraphics[width=0.45\textwidth]{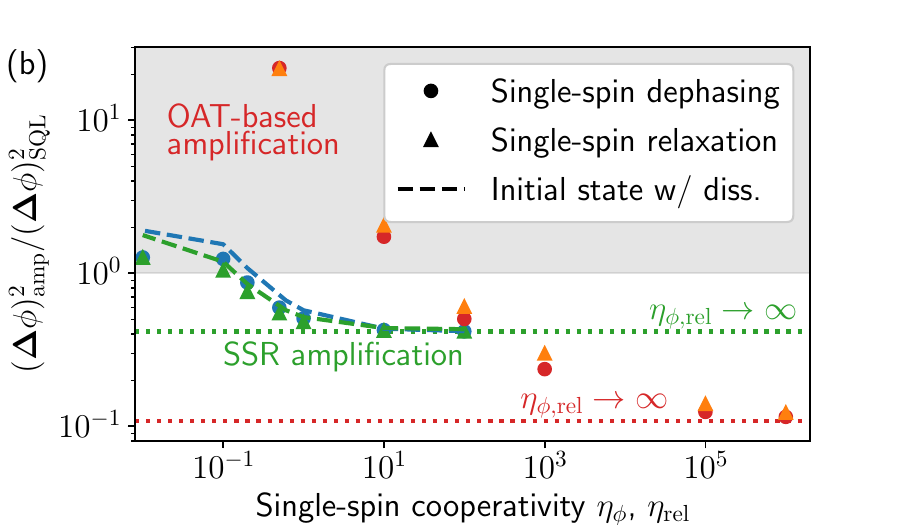}
	\caption{
		Minimum squared estimation error $\Deltaphisqamp$ after amplification [defined in Eq.~\eqref{eqn:EstimationErrorWithAmplification}], relative to the SQL value $\DeltaphisqSQL = 1/N$, (a) as a function of the number of spins for fixed single-spin cooperativities $\etaphi$, $\etarel$ and (b) as a function of the single-spin cooperativity for fixed $N = \FigSsrOatVsEtaValueN$.	
		In both plots, the signal is $\phi = \FigSsrOatVsNValuesignal$ and the detection noise is at the level of the projection noise of a coherent-spin state (CSS), $\Xidetsq = \FigSsrOatVsNValueXidetsq$.
		In the gray shaded region, the estimation error is worse than the SQL. 
		(a) 	The green (blue) markers correspond to squeezed-superradiance (SSR) amplification in the presence of single-spin dephasing (relaxation) with single-spin cooperativity $\etaphi = \Gamma / \gammaphi = \FigSsrOatVsNValueetaphi$ ($\etarel = \Gamma / \gammarel = \FigSsrOatVsNValueetarel$).
		The red (orange) markers show the corresponding minimum $\Deltaphisqamp$ for a unitary amplification scheme using one-axis-twist (OAT) dynamics, which is more than an order of magnitude worse than the SSR result and fails to surpass the SQL.		
		The SSR data points have been obtained from a numerically exact solution of the quantum master equation (QME) shown in Eq.~\eqref{eqn:QME}, and an optimization of the amplification time $t$ and the squeezing parameters $\rini$, $\rdec$.
		The OAT dynamics is implemented using a Tavis-Cummings coupling to a detuned bosonic mode (see \cite{SM}). 
		The solid lines are mean-field theory (MFT) simulations for the same parameters as the markers of the corresponding color (see \cite{SM}). 
		(b) The green and blue (red and orange) markers show the results for SSR-based (OAT-based) amplification in the presence of undesired single-spin dephasing and relaxation, respectively [for SSR, this is modeled by the QME~\eqref{eqn:QME_local_dissipation}]. 
		The dashed lines show the performance of the SSR scheme if single-spin dissipation degrades the preparation of the squeezed initial state (see \cite{SM} for details).
		In both plots, the green (red) dotted line corresponds to the ideal performance of SSR-based (OAT-based) amplification in the absence of undesired single-spin (single-spin and collective) dissipation. 
	}
	\label{fig:ImprovementOverSQL}
\end{figure*}

We consider an ensemble of $N$ identical spin-$1/2$ systems described by the Hamiltonian $\hat{H}_0 = \omega \hat{S}_z$, where $\hat{S}_\alpha = \sum_{j=1}^N \hat{\sigma}_\alpha^j/2$ denotes the $\alpha \in \{x,y,z\}$ components of the collective spin operator and $\hat{\sigma}_\alpha^j$ denotes the Pauli matrices acting on spin $j$. 
Our goal is to estimate a small signal $\phi \to 0$ that causes a nonzero transverse polarization $\cerw{\hat{S}_y}_\mathrm{ini} \propto \phi$, as sketched in Fig.~\ref{fig:Sketch}. 
A general way of generating such a polarized state is to prepare the ensemble in a SSS and to perform a Ramsey measurement where the parameter of interest changes the spin precession frequency. 
The estimation error with which $\phi$ can be inferred from a measurement of $\cerw{\hat{S}_y}_\mathrm{ini}$ depends on the intrinsic projection noise $\DeltaSysqini = \cerw{\hat{S}_y^2}_\mathrm{ini} - \cerw{\hat{S}_y}^2_\mathrm{ini}$ of the state of the ensemble, and additional fluctuations $\Sigmadetsq = \Xidetsq N/4$ caused by the detection process \cite{Degen2017}.
Here, $\Xidetsq$ quantifies the amount of detection noise in units of the projection noise $N/4$ of a simple product state, i.e., a CSS.
Spin squeezing decreases the intrinsic projection noise $\DeltaSysqini$ and is therefore a good strategy to reduce the estimation error if the condition $\Sigmadetsq \ll \DeltaSysqini$ holds. 
In many experimentally relevant spin systems, however, the readout noise is orders of magnitude larger than the intrinsic projection noise, $\Sigmadetsq \gg \DeltaSysqini$.
For instance, the photon shot noise of standard fluorescence detection of nitrogen-vacancy defect centers in diamond has $\Xidetsq \ggg 1$ \cite{Barry2020}.

The detrimental impact of detection noise can be reduced by amplification of the transverse polarization before readout, i.e., $\cerw{\hat{S}_y}_\mathrm{amp} = G \cerw{\hat{S}_y}_\mathrm{ini}$ with a gain factor $G > 1$.
The estimation error then takes the form
\begin{align}
	\Deltaphisqamp
	&= \Deltaphisqproj + \Deltaphisqdet + \Deltaphisqadd \nonumber \\
	&= \frac{\DeltaSysqini + (\Xidetsq/G^2) N/4 + \sigmaaddsq N/4}{\cabs{\partial_\phi \cerw{\hat{S}_y}_\mathrm{ini}}^2} \comma
	\label{eqn:EstimationErrorWithAmplification}
\end{align}
where $\sigmaaddsq$ describes additional quantum fluctuations due to the amplification dynamics, in units of the projection noise of a CSS. 
The estimation error without amplification is obtained by setting $G \to 1$ and $\sigmaaddsq \to 0$. 
Amplification effectively reduces the detection noise, $\Xidetsq \to \Xidetsq/G^2$, and the minimum estimation error is given by the optimal trade-off between projection noise, added noise, and reduced detection noise. 
Amplification protocols that isotropically amplify any transverse polarization have $\sigmaaddsq = \mathcal{O}(1)$ \cite{Koppenhoefer2022} (similar to the corresponding limits for linear bosonic phase-preserving amplifiers \cite{Caves1982,Clerk2010}).
In this case, spin squeezing cannot be used to improve $\Deltaphisqamp$ even if $G^2 \gg \Xidetsq$ because, no matter how small $\Deltaphisqproj$ has been made by the spin squeezing, the unavoidable added noise term $\Deltaphisqadd \geq \sigma_\mathrm{add}^2/N$ is at best at the SQL.
Unitary twist-untwist amplification protocols have $\sigmaaddsq \to 0$ and can therefore surpass the SQL, but they are not robust against undesired dissipation (see below). 
Our goal is thus to find a robust dissipative amplification process with $G \gg 1$ (such that detection noise is suppressed) and $\sigmaaddsq \to 0$ in the $y$ direction (such that an initial spin-squeezed state decreases $\Deltaphisqamp$ below the SQL value $\DeltaphisqSQL = 1/N$).

\prlsection{Squeezed bosonic amplification}

For the simpler problem of reading out a single bosonic mode (with annihilation operator $\hat{a} = (\hat{q}_+ + i \hat{q}_-)/\sqrt{2}$), a dissipative process with these desired properties has recently been demonstrated by Delaney \emph{et al.}~\cite{Delaney2019}:
it amplifies the quadratures $\hat{q}_\pm$ exponentially with the same gain $G_\mathrm{bos}(t) = \exp(\Gamma t/2)$, but adds minimal noise to the $\hat{q}_-$ quadrature: 
\begin{align}
	\frac{\d}{\d t} \hat{q}_{\pm}(t) = \frac{\Gamma}{2} \hat{q}_\pm(t) - \sqrt{\Gamma} \hat{\xi}_\pm(t) \comma
	\label{eqn:BosonicCase:Langevinqpm}
\end{align}
where the fluctuations of $\hat{q}_\pm$ differ exponentially in $\rdec$, $\cerw{\hat{\xi}_\pm(t) \hat{\xi}_\pm(t')} = \delta(t - t')  e^{\pm 2 \rdec}/2$.
In the limit $r \to \infty$, one can thus amplify a signal encoded in a state squeezed along the $\hat{q}_-$ direction without any added noise, $\sigmaaddsq \to 0$. 
The dynamics of Eq.~\eqref{eqn:BosonicCase:Langevinqpm} corresponds to negative damping generated by a Markovian squeezed reservoir, and can be described by the quantum master equation (QME) $\mathrm{d} \hat{\rho}/\mathrm{d} t = \Gamma \mathcal{D}[\hat{\beta}^\dagger] \hat{\rho}$, where $\hat{\beta} = \cosh(\rdec) \hat{a} - \sinh(\rdec) \hat{a}^\dagger$ is a Bogoliubov mode with a squeezing parameter $\rdec \geq 0$ and $\mathcal{D}[\hat{O}]\hat{\rho} = \hat{O} \hat{\rho} \hat{O}^\dagger - \cakomm{\hat{O}^\dagger \hat{O}}{\hat{\rho}}/2$ is a Lindblad dissipator.

\prlsection{Dissipative amplification by squeezed superradiance}

We now ask whether the spin equivalent of Eq.~\eqref{eqn:BosonicCase:Langevinqpm}, generated by the QME ($\hat{S}_\pm = \hat{S}_x \pm i \hat{S}_y$)
\begin{align}
    \frac{\d}{\d t} \hat{\rho} = \Gamma \mathcal{D}[\cosh(\rdec) \hat{S}_- - \sinh(\rdec) \hat{S}_+] \hat{\rho} \comma
    \label{eqn:QME}
\end{align}
has similar properties and could be used for quantum sensing \cite{FN4}. 
Starting from a highly polarized initial state with $\cerw{\hat{S}_z} \approx N/2$, the QME~\eqref{eqn:QME} describes the collective (i.e., superradiant) decay of a spin ensemble driven by broadband squeezed light (with squeezing parameter $\rdec$) \cite{Agarwal1989,Agarwal1990,Agarwal1994}, and several proposals have been made to engineer this type of dynamics in an experimentally more feasible way, 
including driving transitions in multi-level atoms \cite{Kuzmich1997,DallaTorre2013,Borregaard2017}, driving sideband transitions in trapped ion chains, using spin-phonon coupling in optomechanical systems, or coupling spins to superconducting circuits \cite{Groszkowski2022}.
Equation~\eqref{eqn:QME} is a nontrivial generalization of a spin-only model of superradiance (obtained in the limit $r \to 0$ \cite{Dicke1954,Gross1982}) and can be interpreted as superradiant decay to a squeezed environment, hence the name ``squeezed superradiance'' (SSR) \cite{SanchezMunoz2019,GutierrezJauregui2022,FN4}.

For a highly polarized initial state, Eq.~\eqref{eqn:QME} can be mapped onto the bosonic QME $\mathrm{d} \hat{\rho}/\mathrm{d} t = \Gamma \mathcal{D}[\hat{\beta}^\dagger] \hat{\rho}$ to leading order in $1/N$ using a Holstein-Primakoff approximation. 
Thus, at short times and for $e^{2 \rdec} \ll N/2$, exponential amplification and a decoupling of the fluctuations in the $\hat{S}_x,\hat{S}_y$ components are expected.
However, the large gain required to reduce detection noise can only be achieved beyond the regime of applicability of a Holstein-Primakoff approximation,  where the nonlinearity of the spin system can no longer be ignored and the fluctuations of $\hat{S}_x,\hat{S}_y$ as well as the gain can no longer be optimized independently \cite{SM}.
This poses a potential challenge for spin amplification since the nonlinearity of the spin system may now prevent any amplification dynamics with small $\sigmaaddsq$. 
Surprisingly, small $\sigmaaddsq$ can still be achieved, even with moderate levels of squeezing.

\prlsection{Estimation error below the SQL}

We consider a sensing scheme where the system is initialized in a SSS $\hat{\rho}_\mathrm{ss}(\rini)$, which is the steady state of Eq.~\eqref{eqn:QME} with $\rdec$ replaced by $\rini$. 
The spin squeezing of this state, quantified by the Wineland parameter \cite{Wineland1992,Pezze2018}, increases monotonically with increasing $\rini$.
A Ramsey sequence encodes the signal $\phi$ into the transverse $\hat{S}_y$ polarization of the spin ensemble and rotates the state such that it has a large $\hat{S}_z$ polarization.  
The state after these steps can be written as $\hat{\rho}_0(\rini, \phi) = e^{i \phi \hat{S}_x} e^{i \pi \hat{S}_y} \hat{\rho}_\mathrm{ss}(\rini) e^{-i \pi \hat{S}_y} e^{-i \phi \hat{S}_x}$.
Starting from $\hat{\rho}_0(\rini,\phi)$, we numerically integrate Eq.~\eqref{eqn:QME}, extract the gain $G(t) = \cerw{\hat{S}_y(t)} / \cerw{\hat{S}_y(0)}$, and calculate the total estimation error $\Deltaphisqamp$ according to Eq.~\eqref{eqn:EstimationErrorWithAmplification}. 
For each ensemble size $N$ and value of $\Xidetsq$, we optimize the amplification time $t$ and the squeezing parameters $\rini,\rdec$ to minimize the estimation error $\Deltaphisqamp$.

\begin{figure}
	\centering
		\includegraphics[width=0.23\textwidth]{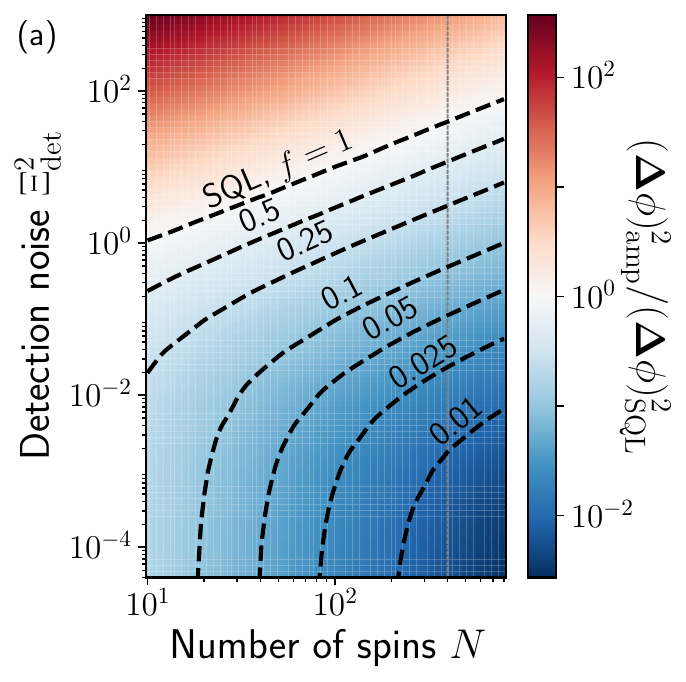}
		\includegraphics[width=0.21\textwidth]{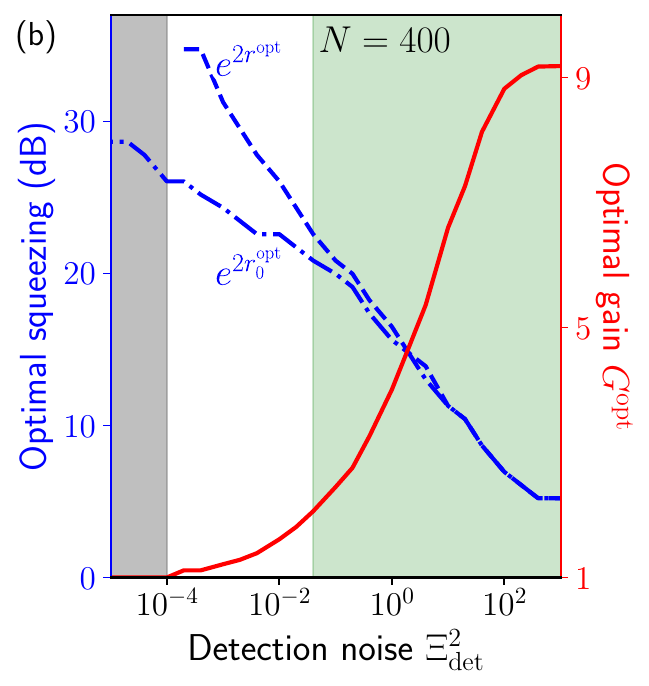}
	\caption{
		(a) Minimum squared estimation error $\Deltaphisqamp$ after amplification, relative to the SQL value $\DeltaphisqSQL = 1/N$, as a function of the number of spins $N$ and the level $\Xidetsq$ of readout noise for $\phi = \FigOptimizedEstimationErrorValuesignal$. 
		The data are linearly interpolated based on an equidistant grid of $\FigOptimizedEstimationErrorValueInterpolationXi$ ($\FigOptimizedEstimationErrorValueInterpolationN$) data points along the $\Xidetsq$ ($N$) direction.
		The contour lines indicate the maximum level of readout noise tolerable if the estimation error should stay the indicated factor $f$ below the SQL. 
		For large $N$, the contour lines scale as $0.13 f^{2.17} N$.
		(b) 	Optimal values of the gain $G$ and squeezing parameter $\rdec$ ($\rini$) of the SSR decay (initial state) for $N=\FigOptimalSqueezingGainValueN$ [indicated by the gray vertical line in (a)]. 
		In the green shaded area on the right, the amplification reduces the estimation error by more than a factor of two compared to no amplification. 
		In the gray shaded area on the left, amplification is not used since it does not improve the estimation error. 
	}
	\label{fig:Fig3}
\end{figure}

As shown in Figs.~\ref{fig:ImprovementOverSQL}(a) and~\ref{fig:Fig3}, SSR decay combined with an initial SSS can surpass the SQL even for significant levels of detection noise, $\Xidetsq \gtrsim 1$. 
Note that the amount of detection noise $\Xidetsq$ that can be tolerated while still providing a sensitivity below the SQL grows linearly with the number of spins for large $N$, as shown by the contour lines in Fig.~\ref{fig:Fig3}(a). 
Unlike in the bosonic case, where the estimation error decreases monotonically with increasing squeezing strength, the spin system reaches the lowest estimation error at finite (and moderate) squeezing parameters $\rdec$ and $\rini$.
In the regime where amplification improves the estimation error by more than a factor of two compared to no amplification [green shaded area in Fig.~\ref{fig:Fig3}(b)], we find that it is optimal to match squeezing parameters, i.e., $\riniopt \approx \rdecopt$. 
Intuitively, this follows from the fact that SSR decay is seeded by the quantum fluctuations of both the squeezed bath and the initial SSS.
Therefore, whenever a large gain is needed, it is best to match the amount of squeezing in the initial SSS to the level of squeezing of SSR \cite{FN2}.
The limiting cases $\Xidetsq \ll 1$ and $\Xidetsq \gg 1$ are discussed in the supplemental material (SM) \cite{SM,FN5}

The estimation error scales as $\Deltaphisqamp \propto 1/N^2$ for $\Xidetsq \ll 1$ and $\Xidetsq \gg 1$, and we numerically find $\Deltaphisqamp \propto 1/N^{3/2}$ for $\Xidetsq \approx 1$ due to the unavoidable presence of added noise in our scheme (see SM \cite{SM}). 
Thus, while SSR amplification allows one to surpass the SQL, it cannot reach the ultimate Heisenberg-limit (HL) scaling.
Unitary amplification strategies, on the other hand, can amplify certain initial states without any added noise and thus seem to provide superior HL-like sensitivity, as shown by the red dotted line in Fig.~\ref{fig:ImprovementOverSQL}(a). 
However, this is no longer the case if undesired dissipative processes are properly taken into account.

\prlsection{Resilience against single-spin dissipation}

We now analyze the impact of unwanted single-spin relaxation (dephasing) at a rate $\gammarel$ ($\gammaphi$) by replacing~Eq.~\eqref{eqn:QME} with 
\begin{align}
	\frac{\d}{\d t} \hat{\rho} 
	&= \Gamma \mathcal{D}[\cosh(\rdec) \hat{S}_- - \sinh(\rdec) \hat{S}_+] \hat{\rho} \nonumber \\
	&+ \gammarel \sum_{j=1}^N \mathcal{D}[ \hat{\sigma}_-^{(j)}] \hat{\rho}
	+ \frac{\gammaphi}{2} \sum_{j=1}^N \mathcal{D}[ \hat{\sigma}_z^{(j)}] \hat{\rho} \fullstop
	\label{eqn:QME_local_dissipation}
\end{align}
Single-spin relaxation (single-spin dephasing) leads to exponential decay of $\hat{S}_x,\hat{S}_y$ (see \cite{SM}) but, starting from a highly polarized initial state $S_z \approx N/2$, the amplification dynamics due to collective decay dominates if the collective cooperativity satisfies $C_\mathrm{rel} \equiv N \Gamma/\gammarel > 1$ ($C_\phi \equiv N \Gamma/\gammaphi > 2$), as shown by the green (blue) data points in Fig.~\ref{fig:ImprovementOverSQL}(a):
For small $N$, single-spin dissipation increases the estimation error, but $\Deltaphisqamp$ quickly recovers the ideal results in the absence of single-spin dissipation (given by the dotted green line) with increasing $N$.

This high level of robustness against single-spin dissipation is in sharp contrast to unitary OAT-based amplification. 
In Fig.~\ref{fig:ImprovementOverSQL}(a), we also show the corresponding results for an amplification scheme based on unitary OAT dynamics \cite{Davis2016}, which has been generated using a strongly detuned Tavis-Cummings interaction between spins and a bosonic mode. 
In this implementation, the desired OAT interaction is accompanied by undesired collective relaxation due to decay of the bosonic mode, as well as single-spin dephasing and relaxation, see \cite{SM} for details.
As shown in Ref.~\cite{Koppenhoefer2022} using exact QME simulations and a mean-field theory (MFT) analysis for large $N$, amplification is only possible if the single-spin cooperativities satisfy $\etaphi \gg \sqrt{N}$ and $\etarel \gg N^{0.9}$. 
Thus, for experimentally realistic values $\etaphi, \etarel = 1/2$ and $N \leq 100$, the OAT amplification scheme misses even the SQL by more than an order of magnitude [red and orange markers in Fig.~\ref{fig:ImprovementOverSQL}(a)], and an estimation error below the SQL even for $\Xidetsq = 1$ is only possible for extremely large $\eta_\phi, \eta_\mathrm{rel}$ [Fig.~\ref{fig:ImprovementOverSQL}(b)]. 
Note that the cooperativity threshold even \emph{increases} for larger $N$ \cite{FN6}. 
In contrast, our scheme readily surpasses the SQL and outperforms unitary OAT protocol in a wide range of experimentally relevant parameters thanks to its much less restrictive requirement of a large \emph{collective} cooperativity.

\prlsection{Conclusion}

We have analyzed an amplification scheme using SSR, i.e., collective decay of an ensemble of $N$ spins in the presence of a squeezed bath. 
Unlike in a related bosonic dissipative amplification scheme, gain and added noise depend on each other due to the nonlinearity of the spin ensemble.
Surprisingly, despite this nonlinearity, it is still possible to achieve an estimation error below the SQL even in the presence of significant levels of detection noise.
Our scheme provides a major practical advantage over related ideas:  
SSR amplification uses the same experimental ingredients as unitary OAT twist-untwist protocols \cite{SM} but,  
being based on collective dissipation, it is much more robust against undesired single-spin dissipation.
All required ingredients to engineer the squeezed bath of the spins have been demonstrated in state-of-the-art experimental platforms \cite{SM,Groszkowski2022}, 
which makes our SSR amplification an interesting candidate to demonstrate quantum sensing with a sensitivity below the SQL in the presence of significant levels of detection noise.

Interesting directions for further research are to combine dissipative and unitary collective dynamics to speed up the generation of spin-squeezed states and to tune the interplay between gain and added noise.
Moreover, nonlinear estimators of the parameter $\phi$ may help to reduce the impact of added noise and increase the sensitivity \cite{Strobel2014}.

\let\oldaddcontentsline\addcontentsline
\renewcommand{\addcontentsline}[3]{}

\begin{acknowledgments}
This work was primarily supported by the DOE Q-NEXT Center (Grant No. DOE 1F-60579). 
We also acknowledge support from the 
Defense Advanced Research Projects Agency (DARPA) Driven and Nonequilibrium Quantum Systems (DRINQS) program (Agreement D18AC00014), and from the Simons Foundation (Grant No. 669487, A. C.).
Finally, this research also used resources of the Oak Ridge Leadership Computing Facility, which is a DOE Office of Science User Facility supported under Contract DE-AC05-00OR22725.
\end{acknowledgments}

\bibliography{citations}

\begin{thebibliography}{64}
\bibitem [{FN2()}]{FN2}%
  \BibitemOpen
  \href@noop {} {}\bibinfo {note} {Since $\hat{\rho}_0(r_0,0)$ is the steady
  state of $\d \hat{\rho}/\d t = \mathcal{D}[\cosh(\rini) \hat{S}_+ -
  \sinh(\rini) \hat{S}_-] \hat{\rho}$, which is related to the
  QME~\eqref{eqn:QME} of SSR decay by a time reversal of the jump operator,
  this result shows that our scheme can be understood as a dissipative
  generalization of unitary twist-untwist protocols.}\BibitemShut {Stop}%
\bibitem [{FN5()}]{FN5}%
  \BibitemOpen
  \href@noop {} {}\bibinfo {note} {In the SM, we also comment on an interesting
  but experimentally challenging regime of an oversqueezed initial state, $e^{2
  \rini} \gg N/2$: this regime also provides both high gain and low added
  noise.}\BibitemShut {Stop}%
\end{thebibliography}
\let\addcontentsline\oldaddcontentsline

\newpage 
\clearpage

\thispagestyle{empty}
\onecolumngrid
\begin{center}
\textbf{\large Supplemental Material for\\\thetitle}
\end{center}

\begin{center}
\theauthors\\
\emph{\theaffiliations}\\
(Dated: \today)
\end{center}

\setcounter{equation}{0}
\setcounter{figure}{0}
\setcounter{table}{0}
\setcounter{page}{1}
\makeatletter
\renewcommand{\theequation}{S\arabic{equation}}
\renewcommand{\thefigure}{S\arabic{figure}}
\renewcommand{\thepage}{S\arabic{page}}
\renewcommand{\citenumfont}[1]{#1}


\tableofcontents

\section{Squeezed superradiance}
\subsection{Additional details on the optimization of $\Deltaphisqamp$}

\begin{figure}[b!]
	\includegraphics[height=5cm]{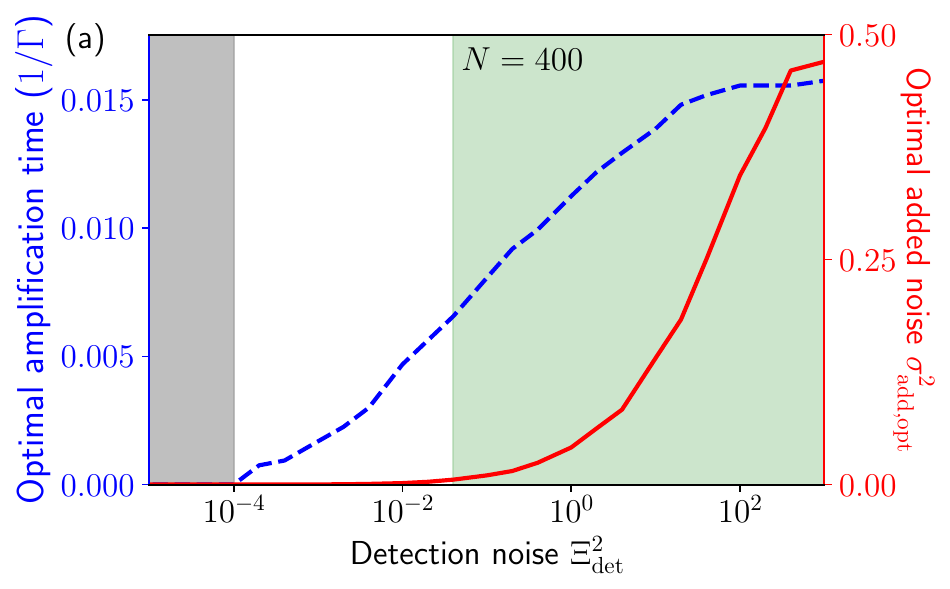}
	\includegraphics[height=5cm]{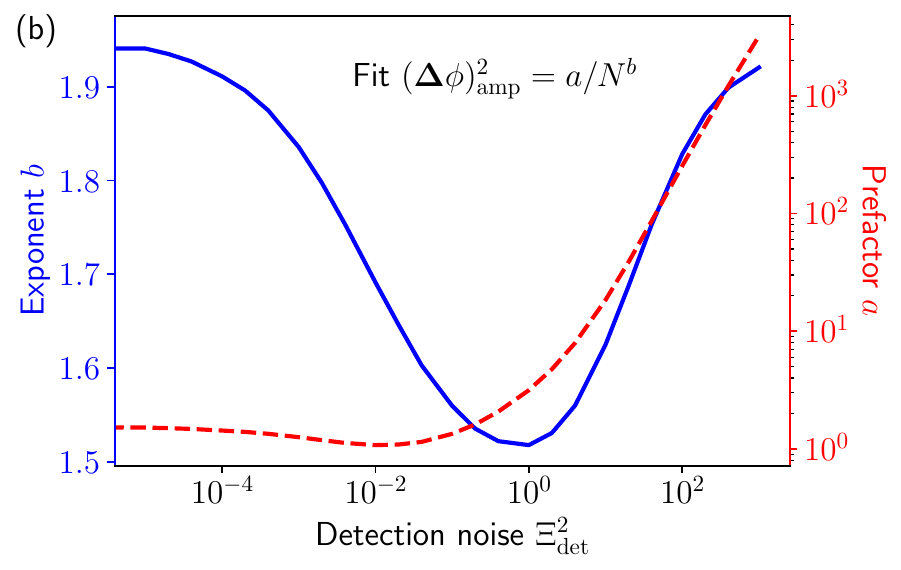}
	\caption{
		Additional plots complementing Fig.~\ref{fig:Fig3} of the main text. 
		(a) Optimal amplification time $\topt$ (blue dashed curve) and optimal added noise $\sigmaaddsqopt$ (solid red curve) as a function of the readout noise $\Xidetsq$ for $N=\FigOptimalTimeAddedNoiseValueN$ spins and $\phi=\FigOptimalTimeAddedNoiseValuesignal$. 
		(b) Prefactor $a$ (dashed red curve) and exponent $b$ (solid blue curve) of the minimum estimation error $\Deltaphisqamp = a / N^b$ as a function of $\Xidetsq$. 
		The data have been obtained by fitting the numerical results shown in Fig.~\ref{fig:Fig3}(a) of the main text in the range $\FigScalingEstimationErrorValueFitRangeLow \leq N \leq \FigScalingEstimationErrorValueFitRangeHigh$ 
	}
	\label{fig:SM:ComplementFig3}
\end{figure}

In this section, we provide additional details on the optimization of the squared estimation error after amplification, $\Deltaphisqamp$, defined in Eq.~\eqref{eqn:EstimationErrorWithAmplification} of the main text. 
Figure~\ref{fig:SM:ComplementFig3}(a) complements Fig.~\ref{fig:Fig3}(b) of the main text and shows the optimal values of the amplification time, $\topt$, and for the added noise, $\sigmaaddsqopt$, for $N=400$ spins. 
Figure~\ref{fig:SM:ComplementFig3}(b) shows the scaling of $\Deltaphisqamp = a / N^b$ with the number of spins, which has been extracted by fitting the data shown in Fig.~\ref{fig:Fig3}(a) of the main text in the range $10 \leq N \leq 400$.

For very small levels of detection noise, $\Xidetsq \to 0$, amplification does not help to improve the estimation error and we find $\topt = \sigmaaddsqopt = 0$ [gray shaded area in Fig.~\ref{fig:SM:ComplementFig3}(a)]. 
In this regime, the estimation error is only determined by the projection noise of the initial state $\hat{\rho}_0(\rini,\phi)$. 
Since squeezed-superradiant (SSR) decay allows one to stabilize a spin-squeezed steady state with a Heisenberg-limit (HL) scaling for $e^{2\rini} \gg N/2$ \cite{Groszkowski2022}, we find $\Deltaphisqamp \to 2 / N^2$.

For larger values of $\Xidetsq$, amplification becomes a useful strategy to reduce the estimation error, and the optimal amplification time $\topt$ grows. 
Practically relevant is the green shaded area in Fig.~\ref{fig:SM:ComplementFig3}(a), where the estimation error is reduced by more than a factor of two compared to the corresponding estimation error without any amplification.

In the limit of extremely large readout noise, $\Xidetsq \gg 0.13 N$, both the intrinsic projection noise and the added noise $\sigmaaddsq = \mathcal{O}(1)$ are negligible compared to the detection noise, such that $\Deltaphisqamp \to \Deltaphisqdet = \Xidetsq/G^2 N$. 
In this regime, the optimal strategy is to maximize the gain irrespective of the added noise, which results in the constant values for $\topt$, $\sigmaaddsqopt$, and $\Gopt \to \Gmax$.
Since $\Gmax \propto \sqrt{N}$, we again find an exponent of $b \approx 2$ but the large prefactor $a \propto \Xidetsq$ shows that we are far from the SQL and this scaling should not be confused with HL-like scaling.
Note that both $\rdecopt$ and $\riniopt$ are nonzero because a small residual amount of squeezing allows one to reduce $\sigmaaddsqopt$ and to increase $\Gopt$ beyond what can be achieved when a coherent-spin state (CSS) is used for sensing combined with ordinary superradiant decay (i.e., $\rdec = 0$).

\begin{figure}
	\centering
	\includegraphics[width=\textwidth]{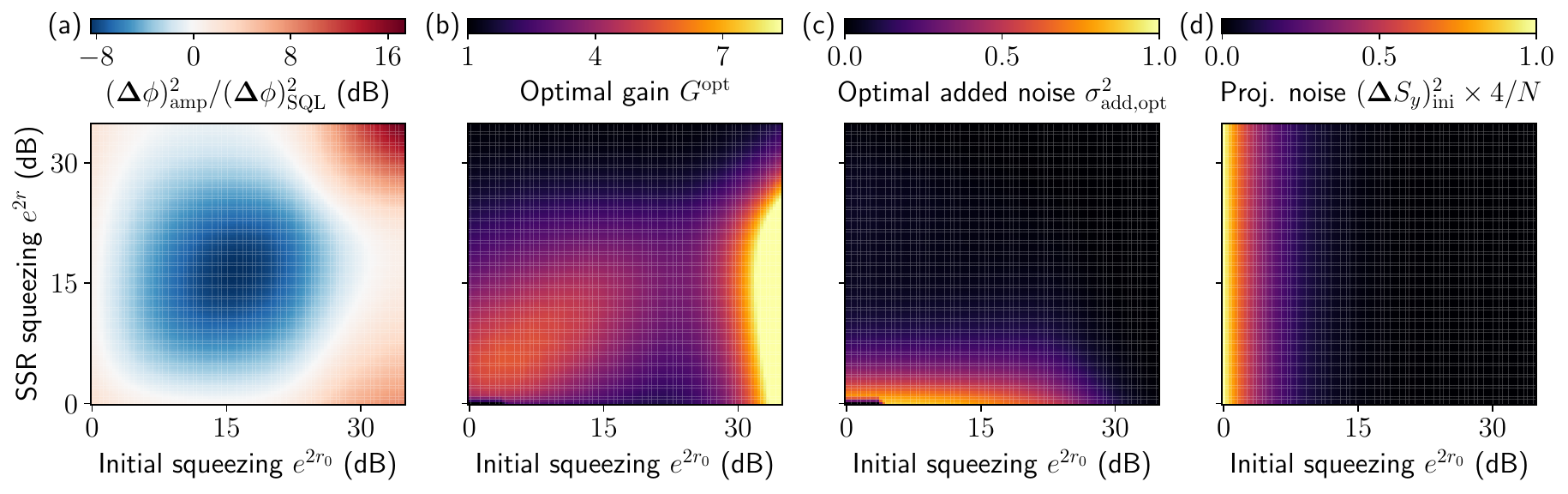}
	\caption{
		Optimal values of (a) the squared estimation error $\Deltaphisqamp$ relative to the SQL value $\DeltaphisqSQL = 1/N$, (b) the gain $\Gopt$, and (c) the added noise $\sigmaaddsqopt$, as well as (d) the fluctuations $\DeltaSysqini$ of the initial state $\hat{\rho}(\rini,0)$ as a function of the squeezing parameters $\rini$ and $\rdec$.
		Parameters are $\Xidetsq = \FigContributionsEstimationErrorValueXidetsq$, $N=\FigContributionsEstimationErrorValueN$, and $\phi = \FigContributionsEstimationErrorValuesignal$. 
		 For each data point, the amplification time $t$ has been optimized to minimize $\Deltaphisqamp$. 
		 The region of high gain for $e^{2 \rini} \gg N/2$ is discussed in Sec.~\ref{sec:SM:AmplificationLargeRini}. 
	}
	\label{fig:SM:2DPlots}
\end{figure}

For a level of readout noise comparable with the projection noise of a CSS, $\Xidetsq \approx 1$, the projection noise, reduced detection noise, and added noise are similar in size and all contribute to the total estimation error.
Due to the nonlinearity of the spin system, none of the three terms can be optimized independently of the others. 
The optimal trade-off correponds to parameters where the optimal gain $\Gopt$ is smaller than the maximally possible gain $\Gmax \propto \sqrt{N}$, which in turn allows one to reduce the added noise below unity, $\sigmaaddsqopt < 1$, as shown in Fig.~\ref{fig:SM:2DPlots}. 
The squeezing parameters $\rini$ and $\rdec$ should be matched (in order to maximize the amplification time and thus the gain) and satisfy 
\begin{align}
	e^{2 \riniopt} \approx e^{2 \rdecopt} \approx 3.1 \frac{N^{0.43}}{\sqrt{\Xidetsq}} \fullstop
\end{align}
Likewise, the optimal gain scales as
\begin{align}
	G_\mathrm{opt} \approx 0.1 e^{2 \rdecopt} (\Xidetsq)^{0.7} \fullstop
\end{align}
As a consequence of the trade-off between gain and added noise, the estimation error scales with an exponent $1.5 \lesssim b \leq 2$, which is below the optimal value $b=2$ that could be achieved in the absence of any added noise, but which is still better than a SQL scaling with $b = 1$. 
Note that the minimum exponent $b \approx 1.5$ at $\Xidetsq = 1$ is related to the asymptotic form $\Xidetsq = 0.13 f^{2.17} N$ of the contour lines in Fig.~\ref{fig:Fig3}(a) of the main text.

\subsection{Single-spin dissipation during the generation of the initial spin-squeezed state}
\label{sec:SM:SSR:LocalDissipation}

In Fig.~\ref{fig:ImprovementOverSQL}(a) of the main text, we have assumed that the system can be initialized in a perfectly spin-squeezed state, which is the steady state of the quantum master equation (QME)~\eqref{eqn:QME} of the main text with $\rdec \to \rini$. 
In practice, single-spin relaxation and single-spin dephasing may be present during the preparation of this initial state, as described by Eq.~\eqref{eqn:QME_local_dissipation} of the main text with $\rdec \to \rini$. 
These undesired dissipative processes allow the system to explore total-angular-momentum subspaces with quantum numbers $J < N/2$ and reduce the amount of spin squeezing of the initial state \cite{Groszkowski2022}.
In Fig.~\ref{fig:ImprovementOverSQL}(b) of the main text, we take these processes into account in the dashed blue and green lines, which are obtained as follows. 

In the case of single-spin relaxation [dashed green curve in Fig.~\ref{fig:ImprovementOverSQL}(b)], we calculate the steady state of Eq.~\eqref{eqn:QME_local_dissipation} of the main text with $\rdec \to \rini$, flip this state to the north pole of the collective Bloch sphere such that we obtain the initial state $\hat{\rho}_0(\rini,\phi)$ defined in the main text,
\begin{align}
	\hat{\rho}_0(\rini, \phi) = e^{i \phi \hat{S}_x} e^{i \pi \hat{S}_y} \hat{\rho}_\mathrm{ss}(\rini) e^{-i \pi \hat{S}_y} e^{-i \phi \hat{S}_x} \comma
	\label{eqn:RhoInitial}
\end{align}
and let it undergo SSR decay for a time $t$ by evolving it using Eq.~\eqref{eqn:QME_local_dissipation} of the main text. 
We then optimize the overall estimation error $\Deltaphisqamp$ over $\rini$, $\rdec$, and the amplification time $t$. 

In the case of single-spin dephasing [dashed blue curve in Fig.~\ref{fig:ImprovementOverSQL}(b)], it is known that the steady state of Eq.~\eqref{eqn:QME_local_dissipation} of the main text has at best $-3\,\mathrm{dB}$ of spin squeezing, but much higher levels of spin squeezing can be achieved at transient times \cite{Groszkowski2022}. 
We therefore initialize the system in a CSS pointing to the south pole of the collective Bloch sphere and evolve the system using Eq.~\eqref{eqn:QME_local_dissipation} with $\rdec \to \rini$ until the Wineland parameter $\xi_\mathrm{R}^2 = N \DeltaSysq/\cerw{\hat{S}_z}^2$ is minimal. 
We then flip this highly spin-squeezed transient state to the north pole of the collective Bloch sphere as described by Eq.~\eqref{eqn:RhoInitial}, and let it undergo SSR decay for a time $t$ generated by evolving it using Eq.~\eqref{eqn:QME_local_dissipation} of the main text.
As usual, we then optimize the overall estimation error over $\rini$, $\rdec$, and the amplification time $t$. 
Using the transient highly spin-squeezed state is crucial, since the steady state of Eq.~\eqref{eqn:QME_local_dissipation} of the main text does not allow one to surpass the SQL.

\subsection{Mean-field theory analysis of squeezed superradiance}
In this section, we analyze SSR decay using mean-field-theory (MFT) to provide additional intuition for the amplification process and to support our numerical results with approximate simulations for much larger ensemble sizes.

\subsubsection{Mean-field equations of motion}

We use a second-order cumulant expansion \cite{Kubo1962,Zens2019,Groszkowski2022} to derive a closed set of differential equations of motion (EoMs) for the first moments $S_\alpha = \cerw{\hat{S}_\alpha}$ and the covariances $C_{\alpha\beta} = \cerw{\cakomm{\hat{S}_\alpha}{\hat{S}_\beta}}/2 - \cerw{\hat{S}_\alpha} \cerw{\hat{S}_\beta}$, where $\alpha,\beta \in \{x,y,z\}$. 
Using the time-dependent gain factors
\begin{align}
	\lambda_\pm(t) = S_z(t) - \frac{1}{2} e^{\pm 2 r} 
\end{align}
defined in the main text,
the QMEs~\eqref{eqn:QME} and~\eqref{eqn:QME_local_dissipation} of the main text generate the following set of MFT EoMs.
\begin{subequations}%
\begin{align}%
	\frac{\d}{\d t} S_x &= \Gamma \left[ C_{xz} + \lambda_+(t) S_x \right] - \frac{\gammarel}{2} S_x - \gammaphi S_x \comma \displaybreak[1]\\
	\frac{\d}{\d t} S_y &= \Gamma \left[ C_{yz} + \lambda_-(t) S_y \right] - \frac{\gammarel}{2} S_y - \gammaphi S_y \comma \displaybreak[1]\\
	\frac{\d}{\d t} S_z &= - \Gamma \left[ C_{xx} + C_{yy} + S_x^2 + S_y^2 + \cosh(2r) S_z \right]  - \gammarel \left( S_z + \frac{N}{2} \right) \comma \displaybreak[1]\\
	\frac{\d}{\d t} C_{xx} &= \Gamma \left[ 2 \lambda_+(t) C_{xx} + e^{+2r} \lambda_-(t) S_z + e^{+2r} C_{zz} + 2 C_{xz} S_x\right] + \gammarel \left( \frac{N}{4} - C_{xx} \right) + \gammaphi \left( \frac{N}{2} - 2 C_{xx} \right) \comma \displaybreak[1]\\
	\frac{\d}{\d t} C_{xy} &= \Gamma \left[ C_{yz} S_x  +  C_{xz} S_y + 2 C_{xy} S_z - \cosh(2r) C_{xy} \right] - \gammarel C_{xy} - 2 \gammaphi C_{xy} \comma \displaybreak[1]\\
	\frac{\d}{\d t} C_{xz} 
		&= \Gamma \left[ \lambda_-(t) C_{xz} - 2 e^{+2r} C_{xz} + \frac{1}{4} S_x + C_{zz} S_x - 2 C_{xx} S_x  - 2 C_{xy} S_y - e^{+2r} S_x S_z \right] \nonumber \\
		&\phantom{=}\ + \gammarel \left( \frac{1}{2} S_x - \frac{3}{2} C_{xz} \right) - \gammaphi C_{xz} \comma \displaybreak[1]\\
	\frac{\d}{\d t} C_{yy} 
		&= \Gamma \left[ 2 \lambda_-(t) C_{yy} + e^{-2r} \lambda_+(t) S_z + e^{-2r} C_{zz} + 2 C_{yz} S_y \right] + \gammarel \left( \frac{N}{4} - C_{yy} \right) + \gammaphi \left( \frac{N}{2} - 2 C_{yy} \right) \comma \displaybreak[1]\\
	\frac{\d}{\d t} C_{yz} 
		&= \Gamma \left[ \lambda_+(t) C_{yz} - 2 e^{-2 r} C_{yz} + \frac{1}{4} S_y + C_{zz} S_y -  2 C_{yy} S_y - 2 C_{xy} S_x - e^{-2r} S_y S_z \right] \nonumber \\
		&\phantom{=}\ + \gammarel \left(\frac{1}{2} S_y - \frac{3}{2} C_{yz} \right) - \gammaphi C_{yz} \comma \displaybreak[1]\\ 	
	\frac{\d}{\d t} C_{zz} 
		&= \Gamma \left[ e^{-2r} C_{yy}  + e^{+2r} C_{xx} - 2 \cosh(2r) C_{zz} + S_z + e^{+2 r} S_x^2 +  e^{-2r} S_y^2 - 4 C_{xz} S_x - 4 C_{yz} S_y \right] \nonumber \\
		&\phantom{=}\ + \gammarel \left(  \frac{N}{2} -  2 C_{zz} + S_z \right) \fullstop
\end{align}%
\label{eqn:SM:MFT}%
\end{subequations}%

\subsubsection{Approximate MFT EoMs and impact of the spin nonlinearity}
\label{sec:MFTandNonlinearity}

To gain intuition for the SSR amplification dynamics, we simplify the MFT EoMs~\eqref{eqn:SM:MFT} further, using the fact that the initial state $\hat{\rho}_0(r_0,\phi)$ defined in Eq.~\eqref{eqn:RhoInitial}
has the following moments and covariances to leading order in $\phi$ (and, for $C_{zz}$, in $\rini$).
\begin{subequations}%
\begin{align}%
	S_x &= 0 \comma &
	S_y &\approx \frac{N}{2} \phi \comma & 
	S_z &\approx \frac{N}{2} \comma \\ 
	C_{xx} &\approx \frac{N}{2} e^{+2 \rini} \comma &
	C_{xy} &= 0 \comma &
	C_{xz} &= 0 \comma \\
	C_{yy} &\approx \frac{N}{2} e^{-2 \rini} \comma & 
	C_{yz} &= - \frac{N}{2} \phi \comma & 
	C_{zz} &\approx N \rini^2 \fullstop 
\end{align}%
\end{subequations}%
These relations allow us to drop all terms of $\mathcal{O}(\phi^2)$ in Eqs.~\eqref{eqn:SM:MFT}.
In addition, we focus only on the collective part of the dynamics, i.e., we ignore local dissipation, $\gammarel = \gammaphi = 0$.

Defining the instantaneous gain factor 
\begin{align}
	\lambda_\pm(t) = S_z(t) - \frac{1}{2} e^{\pm 2 \rdec} \comma 
	\label{eqn:Lambdapm}
\end{align}
we can rewrite the equations of motion for the transverse polarizations in the simple form
\begin{align}
	\frac{\d}{\d t} S_{x,y}(t) &= \Gamma \lambda_{\pm}(t) S_{x,y}(t) \fullstop 
	\label{eqn:MFTSpin:Sxy}
\end{align}
This is very reminiscent of Eq.~\eqref{eqn:BosonicCase:Langevinqpm} of the main text, except that the gain factors in the $\hat{S}_x$ and $\hat{S}_y$ direction are different and time dependent.
Amplification occurs as long as $S_z(t) \geq e^{\pm 2 r}/2$, where the factors $e^{\pm 2 \rdec}/2$ reflect the fact that a two-level system driven by squeezed vacuum noise experiences different decoherence rates in the $\hat{S}_x$ and $\hat{S}_y$ directions \cite{Gardiner1986,Murch2013,Kiilerich2017,Govia2022}.

The EoM for $S_z$ can be derived using the conservation of total angular momentum, $C_{xx} + C_{yy} + C_{zz} + S_z^2 = N/2(N/2+1) + \mathcal{O}(\phi^2)$,
\begin{align}
	\frac{\d}{\d t} S_z(t) = - \Gamma \left[ \frac{N}{2} \left( \frac{N}{2} + 1 \right) - C_{zz}(t) - S_z(t) \Big( S_z(t) - \cosh(2 \rdec) \Big) \right] \comma 
	\label{eqn:MFTSpin:Sz} 
\end{align}
which shows that the quantum fluctuations seeding the superradiant decay [given by the $\cosh(2r)$ term] increase and speed it up for $\rdec > 0$. 
Thus, for large enough $\rdec$, the gain factors $\lambda_\pm(t)$ decay more quickly and the overall gain decreases [unlike in the bosonic case, where $G_\mathrm{bos}(t)$ is independent of $\rdec$].

On the other hand, $\rdec$ should be nonzero to suppress the added noise $\sigmaaddsq$ in the $\hat{S}_y$ direction, such that one can reduce $\Deltaphisqamp$ using a squeezed initial state with $C_{yy} \ll N/4$. 
At first glance, the EoMs for the covariances seem to suggest that $\sigmaaddsq \to 0$ can be reached in the limit $\rdec \to \infty$, similar to the bosonic case: 
\begin{align}
	\frac{\d}{\d t} C_{xx,yy}(t) = \Gamma \left[ 2 \lambda_\pm(t) C_{xx,yy}(t) + e^{\pm 2 \rdec} \lambda_\mp(t) S_z(t) + e^{\pm 2 \rdec} C_{zz}(t) \right] \fullstop
	\label{eqn:MFTSpin:Cxxyy}
\end{align}
The first term describes the expected amplification of the initial covariances, whereas the last two terms describe the added noise and can be exponentially suppressed in the EoM for $C_{yy}$ by increasing $\rdec$. 
However, the last term $\propto C_{zz}$ (which has no bosonic counterpart, see Sec.~\ref{sec:SM:ComparisonSpinBoson}) can still grow large since $C_{zz}$ couples the two covariances $C_{xx}$ and $C_{yy}$, 
\begin{align}
	\frac{\d}{\d t} C_{zz}(t) = \Gamma \left[ e^{+2 \rdec} C_{xx}(t) + e^{-2 \rdec} C_{yy}(t) + S_z(t) - 2 \cosh(2 \rdec) C_{zz}(t) \right] \fullstop
	\label{eqn:MFTSpin:Czz}
\end{align}
Therefore, even if one starts from a SSS, the squeezed covariance $C_{yy}$ will grow because it is driven by $C_{zz}$, which in turn is driven by $e^{+2 \rdec} C_{xx}$.
From this result, one may worry that the nonlinearity prevents amplification with large gain and low added noise in the spin system. 
Surprisingly, as we show in the main text, this is not the case, i.e., large gain and small $\sigmaaddsq$ can be achieved even with moderate levels of squeezing.

\subsubsection{Comparison with the bosonic mean-field equations}
\label{sec:SM:ComparisonSpinBoson}

In this section, we compare the nonlinear MFT equations for $S_{x,y}$ and $C_{xx,yy}$, Eqs.~\eqref{eqn:MFTSpin:Sxy} and~\eqref{eqn:MFTSpin:Cxxyy}, with the corresponding set of linear bosonic MFT equations. 
The Holstein Primakoff approximation allows one to map a highly polarized spin system onto a bosonic mode using the replacements $\hat{S}_+ \to \sqrt{N} \hat{a}$, $\hat{S}_- \to \sqrt{N} \hat{a}^\dagger$, $\hat{S}_z \to N/2$.
In our case, this approximation is possible in the limit $N/2 \gg e^{2 \rdec}, e^{2 \rini}$, where the spin MFT EoMs take the following form:
\begin{subequations}%
\begin{align}%
	\frac{\d}{\d t} S_{x,y} &\approx \frac{\Gamma N}{2} S_{x,y} \comma \\
	\frac{\d}{\d t} C_{xx,yy} &\approx \Gamma N \left( C_{xx,yy} + \frac{N}{4} e^{\pm 2 \rdec} \right) + e^{\pm 2 \rdec} C_{zz} + \mathcal{O}(N) \fullstop 
\end{align}%
\label{eqn:SM:MFT:SpinHPLimit}%
\end{subequations}%
On the other hand, the Holstein-Primakoff approximation maps Eq.~\eqref{eqn:QME} of the main text onto the following QME describing incoherent pumping of a Bogoliubov mode, 
\begin{align}
	\frac{\d}{\d t} \hat{\rho} = N \Gamma \mathcal{D}[ \cosh(r) \hat{a}^\dagger - \sinh(r) \hat{a} ] \hat{\rho} \comma
\end{align} 
where the factor $N$ captures the collective enhancement of the relaxation rate. 
Using the bosonic quadratures $\hat{q}_+ = (\hat{a} + \hat{a}^\dagger)/\sqrt{2}$ and $\hat{q}_- = (\hat{a} - \hat{a}^\dagger)/\sqrt{2} i$ defined in the main text, and using the abbreviations $q_\alpha = \cerw{\hat{q}_\alpha}$ and $C_{\alpha\beta} = \cerw{\cakomm{\hat{q}_\alpha}{\hat{q}_\beta}}/2 - q_\alpha q_\beta$, where $\alpha,\beta \in \{+,-\}$, we can derive the following bosonic MFT equations:
\begin{subequations}%
\begin{align}%
	\frac{\d}{\d t} q_\pm &= \frac{\Gamma N}{2} q_\pm \comma \\
	\frac{\d}{\d t} C_{\pm \pm} &= \Gamma N \left( C_{\pm \pm} + \frac{e^{\pm 2r}}{2} \right) \fullstop
\end{align}%
\label{eqn:SM:MFT:BosonicLimit}%
\end{subequations}%
Since the Holstein Primakoff approximation maps $\hat{S}_{x,y} \to \sqrt{N/2} \hat{q}_\pm$ and $C_{xx,yy} \to (N/2) C_{\pm \pm}$, the two sets of MFT equations~\eqref{eqn:SM:MFT:SpinHPLimit} and~\eqref{eqn:SM:MFT:BosonicLimit} are identical up to the third term $\propto C_{zz}$ in the EoM for $C_{xx,yy}$.

\subsubsection{Details on the numerical simulation of the MFT EoMs}

The MFT results given by the solid green and blue lines in Fig.~\ref{fig:ImprovementOverSQL}(a) of the main text have been obtained as follows. 
We first calculate the steady state of the EoMs~\eqref{eqn:SM:MFT} by numerical integration in the absence of single-spin dissipation, $\gammarel = \gammaphi = 0$, starting from the ground state $S_x = S_y = 0$, $S_z = -N/2$, $C_{xx} = C_{yy} = N/4$ and $C_{xy} = C_{xz} = C_{yz} = C_{zz} =0$. 
The steady state is then rotated to the north pole of the collective Bloch sphere as described by Eq.~\eqref{eqn:RhoInitial}.
With this rotated state, we integrate Eqs.~\eqref{eqn:SM:MFT} again to simulate SSR decay and we optimize over the amplification time $t$ as well as the squeezing parameters $\rini$ and $\rdec$. 

Alternatively, it is also possible to include undesired single-spin dissipation both in the generation of the initial sensing state and in the superradiant amplification step, as described in Sec.~\ref{sec:SM:SSR:LocalDissipation}. 
In this case, the MFT results predict a slighly larger estimation error $\Deltaphisqamp$, but the estimation error is still below the SQL for $N \gtrsim 20$.

\subsection{Amplification for large squeezing parameters}
\label{sec:SM:AmplificationLargeRini}

\begin{figure}
	\centering
	\includegraphics[height=5cm]{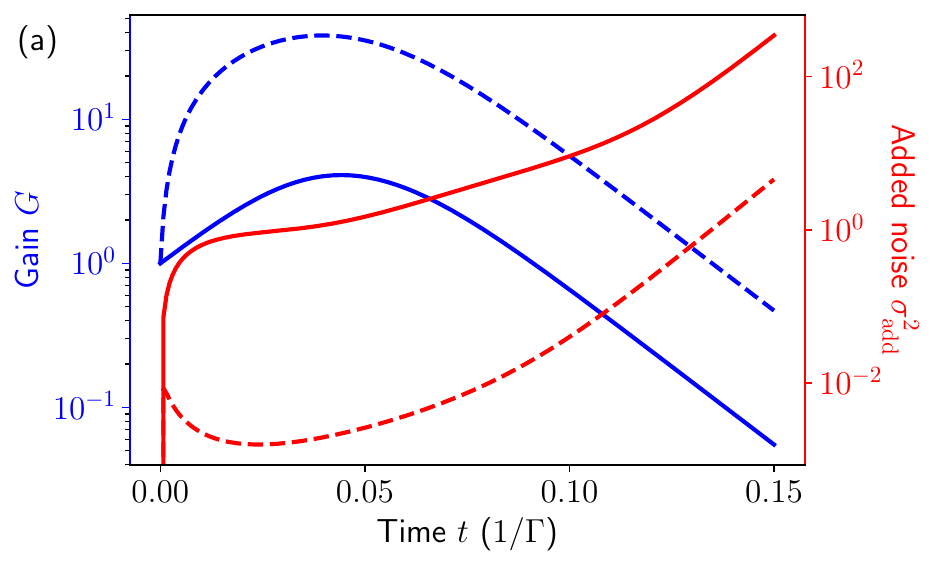}
	\includegraphics[height=5cm]{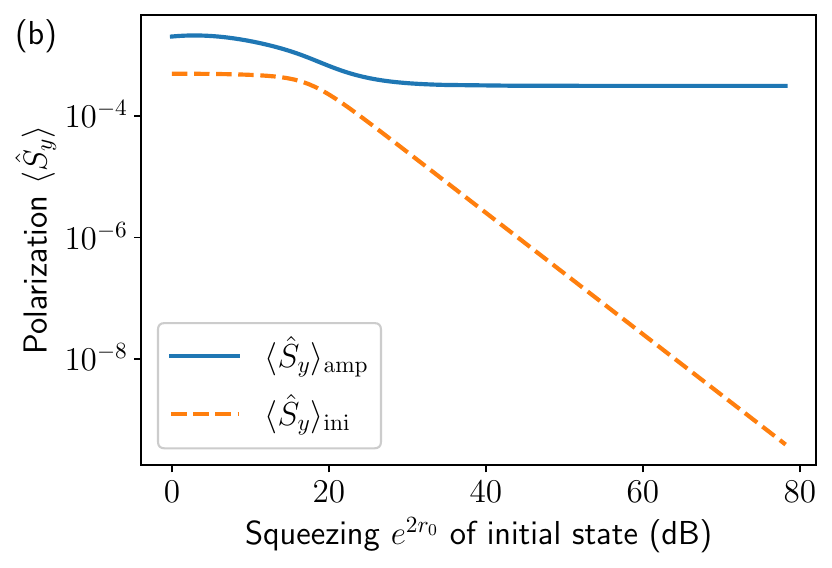}
	\caption{
		Amplification dynamics for large initial squeezing $e^{2 \rini} \gg N/2$, obtained by numerical integration of the QME~\eqref{eqn:QME} of the main text. 
		(a) Gain $G$ (blue curves) and added noise $\sigmaaddsq$ (red curves) as a function of time for $N=\FigGainAndAddedNoiseOversqueezedValueN$, $\phi = \FigGainAndAddedNoiseOversqueezedValuesignal$, and $\rini = \FigGainAndAddedNoiseOversqueezedValueRA$ (solid curves) as well as $\rini = \FigGainAndAddedNoiseOversqueezedValueRB$ (dashed curves). 
		For large initial squeezing, the gain is enhanced and the added noise is reduced. 
		(b) Initial polarization $\cerw{\hat{S}_y}_\mathrm{ini}$ (dashed curve) and maximally amplified polarization $\cerw{\hat{S}_y}_\mathrm{amp}$ (solid curve) as a function of the squeezing $e^{2 \rini}$ of the initial state. 
		The enhanced gain for $e^{2 \rini} \gg N/2$ competes with an exponential suppression of the initial polarization $\cerw{\hat{S}_y}_\mathrm{ini}$, such that the amplified polarization $\cerw{\hat{S}_y}_\mathrm{amp}$ is only moderate and comparable to the \emph{initial} polarization $\cerw{\hat{S}_y}_\mathrm{ini}$ in the opposite regime $e^{2 \rini} \ll N/2$. 
	}
	\label{fig:SM:AmplificationLargeRini}
\end{figure}

Besides the amplification dynamics discussed in the main text, where the squeezing parameters satisfy $e^{2 \rini}, e^{2 \rdec} \ll N/2$, the SSR decay also amplifies highly squeezed initial states with $e^{2 \rini} \gg N/2$. 
In this regime, the initial state $\hat{\rho}_0(\rini, \phi)$ [defined in Eq.~\eqref{eqn:RhoInitial}] resembles a Dicke state $\ket{m_x = 0}$ in the $\hat{S}_x$ basis: 
It has maximum fluctuations in the $\hat{S}_x$ direction, $C_{xx} = N^2/8$, exponentially small fluctuations in the $\hat{S}_y$ direction, $C_{yy} = e^{- 4 \rini} N^2/8$, and an exponentially small net polarization, 
\begin{align}
	\cerw{\hat{S}_z} &\approx \frac{N^2}{4} e^{-2 \rini} \cos(\phi) \comma &
	\cerw{\hat{S}_y} &\approx \frac{N^2}{4} e^{-2 \rini} \sin(\phi) \comma & 
	\cerw{\hat{S}_x} &= 0 \fullstop
	\label{eqn:SM:HighSqueezingAmplification:Polarization}
\end{align} 
SSR decay of this state amplifies the small $\hat{S}_y$ polarization (for $\rdec = 0$) by a gain factor
\begin{align}
	G \approx \frac{e^{2 \rini}}{N^{0.9}} \comma
	\label{eqn:SM:HighSqueezingAmplification:Gain}
\end{align}
which can be exponentially larger than the maximum possible gain $\Gmax \propto \sqrt{N}$ found in the regime $e^{2 \rini}, e^{2 \rdec} \lesssim N/2$ discussed in the main text. 
A comparison between the gain $G(t)$ as a function of the amplification time $t$ in the regimes $e^{2 \rini} \gg N/2$ and $e^{2 \rini} \ll N/2$ is shown in Fig.~\ref{fig:SM:AmplificationLargeRini}(a). 
However, it is impractical to use the regime $e^{2 \rini} \gg N/2$ for spin amplification (despite its nominally extremely large gain) for the following reasons. 
\begin{enumerate}
	\item The required level of squeezing, $e^{2 \rini} \gg N/2$, is high even for small spin ensembles, which makes the preparation of $\hat{\rho}_0(\rini,\phi)$ experimentally very challenging:
		Even $N=200$ spins require more than $20\,\mathrm{dB}$ of squeezing, which exceeds the levels of squeezing that have been demonstrated in experimental platforms that could potentially implement SSR decay via reservoir engineering \cite{Kienzler2015,Dassonneville2021,Groszkowski2022}, and which has only been demonstrated in much larger spin ensembles in optical clocks \cite{Hosten2016}. 
	\item Even if one could generate the required levels of squeezing, the magnitude of the signal \emph{after} amplification in the regime $e^{2 \rini} \gg N/2$ will only be comparable to the magnitude of the signal \emph{before} amplification in the regime $e^{2 \rini} \lesssim N/2$:
		Combining Eqs.~\eqref{eqn:SM:HighSqueezingAmplification:Polarization} and~\eqref{eqn:SM:HighSqueezingAmplification:Gain}, we find that the exponential enhancement of the gain factor cancels with the exponential suppression of the initial polarization, yielding a net polarization \emph{after} amplification of 
		\begin{align}
			\cerw{\hat{S}_y}_\mathrm{amp} \approx \frac{N}{4} \phi \comma
		\end{align}
		which is of the same order of magnitude as the polarization $\cerw{\hat{S}_y}_\mathrm{ini} \approx N \phi/2$ of the initial state \emph{before} amplification in the regime $e^{2 \rini} \lesssim N/2$ [see Fig.~\ref{fig:SM:AmplificationLargeRini}(b)]. 
		Moreover, for $e^{2 \rini} \lesssim N/2$, amplification will further enhance $\cerw{\hat{S}_y}_\mathrm{ini}$ by $\Gmax \propto \sqrt{N}$, which makes the regime $e^{2 \rini} \lesssim N/2$ experimentally more attractive since a larger amplified polarization is easier to read out. 
\end{enumerate}

\subsection{Experimental implementation}
\label{sec:SM:ExperimentalImplementation}

In this section, we provide a more detailed discussion of the experimental implementation and feasibility of our spin-amplification protocol.

The collective SSR decay given by Eq.~\eqref{eqn:QME} of the main text can be implemented in various ways:
The most direct (but experimentally hard) approach is to illuminate an ensemble of effective two-level systems by broadband squeezed light \cite{Agarwal1989,Agarwal1990,Agarwal1994,Kuzmich1997,GutierrezJauregui2022}. 
In search of an experimentally more feasible method, Raman schemes in multi-level systems have been proposed \cite{DallaTorre2013,Borregaard2017}. 
Alternatively, collective SSR decay can be implemented by coupling an ensemble of effective two-level systems to an auxiliary bosonic mode (for instance, a cavity mode) with an engineered dissipative environment that relaxes the bosonic mode to a squeezed state \cite{Groszkowski2022}. 
This approach is very generic and allows one to generate SSR decay dynamics in very different platforms such as trapped-ion setups, solid-state spins coupled to an optomechanical crystal, and superconducting microwave cavities.
As discussed in Ref.~\onlinecite{Groszkowski2022}, the individual building blocks to implement the reservoir-engineering process have been demonstrated in each of these platforms, and there are no fundamental obstacles to combining and generate reservoir-engineered SSR decay in a state-of-the-art experiment.

The collective SSR decay term in Eq.~\eqref{eqn:QME} of the main text must dominate over non-collective effects, such as the single-spin relaxation and dephasing processes shown in Eq.~\eqref{eqn:QME_local_dissipation} of the main text, inhomogeneous broadening of the spin transition frequencies, and inhomogeneities in the coupling strength of each spin to the auxiliary mode. 
We stress that this competition between collective and non-collective effects is a generic feature of every sensing protocol using collective dynamics, i.e., it also applies to unitary protocols like OAT. 
Focusing on the reservoir-engineering approach to generate SSR decay, our protocol uses the same spin-cavity coupling as OAT experiments, which have been demonstrated in large spin ensembles \cite{Leroux2010,Hosten2016,Braverman2019,PedrozoPenafiel2020,Colombo2022}.
Hosten \emph{et al.} \cite{Hosten2016} explicitly investigated the impact of coupling inhomogeneities in their experiment and showed that it led only to a $4\%$ reduction in fringe visibility.
They also checked the impact of inhomogeneous broadening but observed no significant degradation of the performance.
In an earlier OAT experiment, inhomogeneous broadening led only to a reduction of the effective number of spins in the ensemble \cite{Leroux2010} but did not disrupt the collective nature of the dynamics.

Moreover, superradiant and subradiant phenomena have already been observed in solid-state platforms \cite{Angerer2018}, spin ensembles coupled to optical cavities \cite{Bohnet2012} as well as directional waveguides \cite{Liedl2023}, and in dense atomic clouds \cite{Ferioli2021}, despite the presence of competing non-collective dynamics.
The coherence of the superradiant laser demonstrated by Bohnet \emph{et al.} \cite{Bohnet2012} stems from the collective decay of the spin ensemble, which led to a linewidth-narrowing by a factor of 10 below the one expected from the competing inhomogeneous broadening in the setup. 
Note that the SSR decay is compatible with the dynamical decoupling protocol introduced in Ref.~\onlinecite{Groszkowski2022}, which allows one to further reduce the detrimental impact of inhomogeneous broadening.
The same protocol can also be used to rapidly switch off the decay rate of SSR decay in order to interrupt the decay dynamics at the point of maximum gain.

In summary, all required experimental ingredients for spin amplification using SSR have already been demonstrated in several experiments, thus making our amplification protocol realizable in state-of-the-art quantum sensing platforms.

\section{Implementation of OAT dynamics using a Tavis-Cummings interaction}
\subsection{Effective quantum master equation}

In this section, we disucss the effective QME for the one-axis-twist (OAT) amplification scheme considered in the main text. 
We consider an ensemble of $N$ spin-$1/2$ systems coupled to a bosonic mode with a Tavis-Cummings Hamiltonian
\begin{align}
	\hat{H}_\mathrm{TC} = \oma \hat{a}^\dagger \hat{a} + \oms \hat{S}_z + g \left( \hat{S}_- \hat{a}^\dagger + \hat{S}_+ \hat{a} \right) \fullstop
\end{align}
Each spin undergoes single-spin dephasing (relaxation) at a rate $\gammaphi$ ($\gammarel$) and the bosonic mode is damped at a rate $\kappa$, which is described by the QME
\begin{align}
	\frac{\d}{\d t} \hat{\rho} = - i \komm{\hat{H}}{\hat{\rho}} + \kappa \mathcal{D}[\hat{a}] \hat{\rho} + \gammarel \sum_{j=1}^N \mathcal{D}[\hat{\sigma}_-^j] \hat{\rho} + \frac{\gammaphi}{2} \sum_{j=1}^N \mathcal{D}[\hat{\sigma}_z^j] \hat{\rho} \fullstop
\end{align}
In the limit of a large detuning between the spins and the bosonic mode, $\abs{\Delta} = \abs{\oma - \oms} \gg g$, one can adiabatically eliminate the bosonic mode using a Schrieffer-Wolff transformation.
In a frame rotating at $\oms$, we find the QME
\begin{align}
	\frac{\d}{\d t} \hat{\rho} 
		= \mathcal{L}_\chi \hat{\rho} 
		= - i \komm{\chi \hat{S}_z^2}{\hat{\rho}} + \Gammacoll \mathcal{D}[\hat{S}_-] \hat{\rho} + \gammarel \sum_{j=1}^N \mathcal{D}[\hat{\sigma}_-^j] \hat{\rho} + \frac{\gammaphi}{2} \sum_{j=1}^N \mathcal{D}[\hat{\sigma}_z^j] \hat{\rho} \comma
	\label{eqn:SM:QME_OAT}
\end{align}
where we defined the OAT strength $\chi = g^2/\Delta$ and the collective decay rate $\Gammacoll = \chi \kappa/\Delta$. 
The decay of the bosonic mode generates collective Purcell decay of the spin ensemble. 
If there was no single-spin dissipation, $\gammarel = \gammaphi = 0$, this undesired Purcell decay could be suppressed by increasing the detuning $\Delta$. 
However, this is not feasible in the presence of single-spin dissipation, because a large detuning would also suppress the desired OAT strength $\chi$ such that single-spin dissipation eventually becomes dominant. 
Therefore, there is an optimal finite ratio $\Delta/\kappa$, which minimizes the detrimental impact of Purcell decay and single-spin dissipation on the OAT dynamics, and maximizes the achievable gain.

As discussed in Ref.~\cite{Koppenhoefer2022}, a crucial ingredient of the OAT-based amplification scheme proposed by Davis \emph{et al.} \cite{Davis2016} is that $\cerw{\hat{S}_z}$ is a constant of motion of the OAT dynamics. 
Starting from an initial state with large $\hat{S}_x$ polarization, the signal is encoded in the $\hat{S}_z$ polarization, $\cerw{\hat{S}_z} \propto \phi$, which ideally remains constant during the subsequent OAT dynamics and causes a linear growth of $\cerw{\hat{S}_y}$ proportional to $\phi$. 
In practice, however, both the Purcell-decay term and the single-spin-relaxation term in Eq.~\eqref{eqn:SM:QME_OAT} break the conservation of $\hat{S}_z$. 
The decay of $\hat{S}_z$ during the amplification process causes an additional background which has to be subtracted to isolate the gain dynamics.
The gain has therefore the form
\begin{align}
	G_\mathrm{sub}^\mathrm{OAT}(t) = \lim_{\phi \to 0} \frac{\cerw{\hat{S}_y(t,\phi)} - \cerw{\hat{S}_y(t,0)}}{\frac{N \phi}{2}} \comma
	\label{eqn:SM:GOAT}
\end{align}
where $\cerw{\hat{S}_y(t,0)}$ denotes the time-dependent value of $\cerw{\hat{S}_y}$ in the absence of any signal, $\phi = 0$. 
Numerical studies showed that $G_\mathrm{sub}^\mathrm{OAT}$ is highly reduced from its ideal value $\sqrt{N/e}$ unless the single-spin cooperativities satisfy $\etaphi = 4 g^2/\kappa \gamma_\phi \gg \sqrt{N}$ and $\etarel = 4 g^2/\kappa \gamma_\mathrm{rel} \gg N^{0.9}$ \cite{Koppenhoefer2022}.

\subsection{Numerical minimization of the estimation error}
\label{sec:SM:OAT:NumericalAnalysis}

The numerical results shown by the red and orange markers in Fig.~\ref{fig:ImprovementOverSQL} of the main text have been obtained from Eq.~\eqref{eqn:SM:QME_OAT} as follows. 
The system is initialized in a CSS $\hat{\rho}_0^\mathrm{OAT}$  pointing along the positive $\hat{S}_x$ direction. 
We evolve $\hat{\rho}_0^\mathrm{OAT}$ for a time $t_\mathrm{sqz}$ by numerically integrating Eq.~\eqref{eqn:SM:QME_OAT} to turn the state into an (over)squeezed state $\hat{\rho}_1^\mathrm{OAT} = e^{\mathcal{L}_\chi t_\mathrm{sqz}} (\hat{\rho}_0^\mathrm{OAT})$. 
Next, the signal is applied by rotating $\hat{\rho}_1^\mathrm{OAT}$ about the $\hat{S}_y$ axis, 
\begin{align}
	\hat{\rho}_2^\mathrm{OAT}(\phi) = e^{-i \phi \hat{S}_y} \hat{\rho}_1^\mathrm{OAT} e^{+i \phi \hat{S}_y} \fullstop
\end{align}
We then undo the squeezing operation by numerically integrating the QME~\eqref{eqn:SM:QME_OAT} with $\chi \to -\chi$ for a time $t_\mathrm{unsqz} = t_\mathrm{sqz}$ starting from $\hat{\rho}_2^\mathrm{OAT}(\phi)$, 
\begin{align}
	\hat{\rho}_\mathrm{final}^\mathrm{OAT}(\phi) = e^{\mathcal{L}_{-\chi} t_\mathrm{unsqz}}[\hat{\rho}_2^\mathrm{OAT}(\phi)] \fullstop
\end{align}
Finally, the gain is calculated using~\eqref{eqn:SM:GOAT} with $\cerw{\hat{S}_y(t,\phi)} = \Tr [ \hat{S}_y \hat{\rho}_\mathrm{final}^\mathrm{OAT}(\phi)]$, and the estimation error in the presence of readout noise is obtained from
\begin{align}
	\Deltaphisqamp = \frac{\DeltaSysqfin + \Xidetsq N/4}{\abs{\frac{N}{2} G_\mathrm{sub}^\mathrm{OAT}}^2} \comma
\end{align}
where $\DeltaSysqfin = \Tr[ \hat{S}_y^2 \hat{\rho}_\mathrm{final}^\mathrm{OAT}(\phi)] - \Tr[ \hat{S}_y \hat{\rho}_\mathrm{final}^\mathrm{OAT}(\phi)]^2$.
For a given single-spin cooperativity $\eta_k = 4 g^2/\kappa \gamma_k$ with $k \in \{\phi, \mathrm{rel}\}$, we optimize the estimation error over the (un)twist time $t_\mathrm{sqz}=t_\mathrm{unsqz}$ and the ratio $\kappa/\Delta$ determining the relative strength of Purcell decay and unitary OAT dynamics.
We focus on the usual case where the twist and untwist times are chosen to be identical \cite{Davis2016}. 
Anders \emph{et al.} \cite{Anders2018} showed that a separate optimization of $t_\mathrm{sqz}$ and $t_\mathrm{unsqz}$ leads only to very small improvements.

In the absence of any undesired dissipation, $\Gammacoll = \gammarel = \gammaphi = 0$, the unitary OAT amplification returns to a CSS state, $\DeltaSysqfin = (\bfDelta S_y)^2_0 = N/4$ and the gain approaches $G_\mathrm{sub,ideal}^\mathrm{OAT} = \sqrt{N/e}$ \cite{Davis2016}. 
Under these idealized conditions, one thus finds 
\begin{align}
	\Deltaphisqampideal = \frac{e (1 + \Xidetsq)}{N^2} \comma
\end{align}
which is shown by the dotted red lines in Fig.~\ref{fig:ImprovementOverSQL}.
From this result, we find that the maximum tolerable level of readout noise such that the estimation error a factor of $f$ still below the SQL is achieved also scales linearly with $N$ for the OAT scheme, $\Xi^2_\mathrm{det,th} =  f N/e - 1$.

\subsection{Mean-field equations}

We support the numerical analysis outlined in Sec.~\ref{sec:SM:OAT:NumericalAnalysis} using MFT simulations, which are shown by the solid red and orange lines in Fig.~\ref{fig:ImprovementOverSQL}(a) of the main text. 
Using a second-order cumulant expansion \cite{Kubo1962,Zens2019,Groszkowski2022}, one can derive the following MFT EoMs from the QME~\eqref{eqn:SM:QME_OAT}.
\begin{subequations}%
\begin{align}%
	\frac{\d}{\d t} S_x 
		&= - 2 \chi (C_{yz}  +  S_y S_z) +  \Gammacoll \left( C_{xz}  - \frac{1}{2} S_x  +  S_x S_z \right) - \gammaphi S_x  -  \frac{\gammarel}{2} S_x \comma \\
	\frac{\d}{\d t} S_y
		&= + 2 \chi (C_{xz}  +  S_x S_z) +  \Gammacoll \left( C_{yz}  - \frac{1}{2} S_y  +  S_y S_z \right) - \gammaphi S_y  -  \frac{\gammarel}{2} S_y \comma \\
	\frac{\d}{\d t} S_z 
		&= - \Gammacoll \left( C_{xx}  +  C_{yy}  +  S_x^2  +  S_y^2  +  S_z \right)  - \gammarel \left( \frac{N}{2}  +  S_z \right) \comma \displaybreak[1]\\
	\frac{\d}{\d t} C_{xx} 
		&= -  4 \chi \left( C_{xz} S_y  + C_{xy} S_z \right) + \Gammacoll \left( C_{zz} - C_{xx} + 2 C_{xz} S_x + 2 C_{xx} S_z - \frac{1}{2} S_z + S_z^2   \right) \nonumber \\
		&\phantom{=}\ +  \gammaphi \left( \frac{N}{2} - 2 C_{xx} \right) + \gammarel \left( \frac{N}{4} - C_{xx} \right) \comma \\
	\frac{\d}{\d t} C_{xy}
		&=  -  2 \chi \left( C_{yz} S_y - C_{xz} S_x - C_{xx} S_z  + C_{yy} S_z \right) + \Gammacoll \left( C_{yz} S_x  +  C_{xz} S_y - C_{xy} + 2 C_{xy} S_z \right) - 2 \gammaphi C_{xy} - \gammarel  C_{xy} \comma \\
	\frac{\d}{\d t} C_{xz}
		&= + \chi \left( \frac{1}{2} S_y -  2 C_{zz} S_y -  2 C_{yz} S_z  \right) + \Gammacoll \left( -  2 C_{xy} S_y + \frac{1}{4} S_x -  2 C_{xx} S_x + C_{zz} S_x - S_x S_z  - \frac{5}{2} C_{xz} + C_{xz} S_z \right) \nonumber \\
		&\phantom{=}\ - \gammaphi C_{xz} + \gammarel \left( \frac{1}{2} S_x - \frac{3}{2} C_{xz} \right) \comma \displaybreak[1]\\
	\frac{\d}{\d t} C_{yy}
		&=  +  4 \chi \left( C_{yz} S_x + C_{xy} S_z \right) + \Gammacoll \left( C_{zz} - C_{yy} +  2 C_{yz} S_y + 2 C_{yy} S_z - \frac{1}{2} S_z + S_z^2 \right) \nonumber \\
		&\phantom{=}\ +  \gammaphi \left( \frac{N}{2} - 2 C_{yy} \right) + \gammarel \left( \frac{N}{4} - C_{yy} \right) \comma \\
	\frac{\d}{\d t} C_{yz}
		&= - \chi \left( \frac{1}{2} S_x - 2 C_{zz} S_x - 2 C_{xz} S_z \right) + \Gammacoll \left( - 2 C_{xy} S_x + \frac{1}{4} S_y - 2 C_{yy} S_y + C_{zz} S_y -  S_y S_z - \frac{5}{2} C_{yz} + C_{yz} S_z \right) \nonumber \\
		&\phantom{=}\ - \gammaphi C_{yz} + \gammarel \left( \frac{1}{2} S_y - \frac{3}{2} C_{yz} \right) \comma \\
	\frac{\d}{\d t} C_{zz}
		&= \Gammacoll \left( C_{xx} +  C_{yy} -  2 C_{zz}  +  S_x^2  +  S_y^2  +  S_z  -  4 C_{xz} S_x  -  4 C_{yz} S_y  \right) + \gammarel \left( \frac{N}{2} + S_z - 2 C_{zz} \right) \fullstop
\end{align}%
\label{eqn:SM:MFT_OAT}%
\end{subequations}%
Using these equations, we repeat the procedure outlined in Sec.~\ref{sec:SM:OAT:NumericalAnalysis} to calculate $\Deltaphisqamp$, and we optimize over the (un)twist time $t_\mathrm{sqz} = t_\mathrm{unsqz}$ as well as the ratio $\kappa/\Delta$.

\end{document}